\documentclass[a4paper, amsfonts, amssymb, amsmath, reprint, showkeys, nofootinbib,superscriptaddress, twoside,aps,prd,longbibliography]{revtex4-1}
\usepackage[english]{babel}
\usepackage[utf8]{inputenc}
\usepackage[colorinlistoftodos, color=green!40, prependcaption]{todonotes}
\usepackage{amsthm}
\usepackage{mathtools}
\usepackage{physics}
\usepackage{xcolor}
\usepackage{graphicx}
\usepackage[left=23mm,right=13mm,top=35mm,columnsep=15pt]{geometry} 
\usepackage{adjustbox}
\usepackage{placeins}
\usepackage[T1]{fontenc}
\usepackage{lipsum}
\usepackage{csquotes}
\usepackage{hyperref}
\usepackage{soul}

\begin{document}
    \title{Discrimination of coherent and incoherent cathodoluminescence using temporal photon correlations}

\author{Michael Scheucher}
    \email[]{michael.scheucher@tuwien.ac.at}
    \affiliation{Vienna Center for Quantum Science and Technology, Atominstitut, TU Wien, Vienna, Austria}
    \affiliation{IQOQI Vienna, Austrian Academy of Sciences, Vienna, Austria}
\author{Thomas Schachinger}
    \affiliation{University Service Centre for Transmission Electron Microscopy (USTEM), TU Wien, Vienna, Austria}
    \affiliation{Institute of Solid State Physics, TU Wien, Vienna, Austria}
\author{Thomas Spielauer}
    \affiliation{Vienna Center for Quantum Science and Technology, Atominstitut, TU Wien, Vienna, Austria}
\author{Michael St\"oger-Pollach}
    \affiliation{University Service Centre for Transmission Electron Microscopy (USTEM), TU Wien, Vienna, Austria}
    \affiliation{Institute of Solid State Physics, TU Wien, Vienna, Austria}
\author{Philipp Haslinger}
  \email[]{philipp.haslinger@tuwien.ac.at}
    \affiliation{Vienna Center for Quantum Science and Technology, Atominstitut, TU Wien, Vienna, Austria}

\date{\today} 

\begin{abstract}
We present a method to separate coherent and incoherent contributions of cathodoluminescence (CL) by using a time-resolved coincidence detection scheme. For a proof-of-concept experiment, we generate CL by irradiating an optical multimode fiber with relativistic electrons in a transmission electron microscope. A temporal analysis of the CL reveals a large peak in coincidence counts for small time delays, also known as photon bunching. Additional measurements allow us to attribute the bunching peak to the temporal correlations of coherent CL (Cherenkov radiation) created by individual electrons. Thereby, we show that coincidence measurements can be employed to discriminate coherent from incoherent CL and to quantify their contribution to the detected CL signal. This method provides additional information for the correct interpretation of CL, which is essential for material characterization. Furthermore, it might facilitate the study of coherent electron-matter interaction.\\
\end{abstract}


\maketitle

\newpage
\section{Introduction}

In recent years, cathodoluminescence (CL) developed to be a powerful tool to characterize optical properties of materials with subwavelength spatial resolution \cite{Kociak2017,Coenen2017}. In general, CL is light in the optical spectral range emitted from a material under electron irradiation. Various different physical processes can lead to CL \cite{GarciaDeAbajo2010}, which are divided into two main classes: incoherent and coherent CL.
The incoherent form of emission involves spectroscopic/quantum mechanical transitions, which are excited by the incoming electron beam and subsequently decay by emitting a photon. Thus, there is no phase relation between the emitted light and the excitation, and the light is typically unpolarized and isotropic.
 Prominent examples for incoherent CL are band edge emission or emission from defect states \cite{GarciaDeAbajo2010}.
The second class of CL, referred to as coherent CL, arises directly from the electric polarization of the sample due to the time-varying electromagnetic near-fields of the passing electron. As a consequence, the emitted light is strongly polarized and its phase is linked to the incoming electron \cite{GarciaDeAbajo2010}. 
Coherent CL includes e.g. transition radiation \cite{Yamamoto2001,Stoger-Pollach2017}, 
 the Smith-Purcell effect \cite{Smith1953}, radiative plasmon decay \cite{Vesseur2007} and  Cherenkov radiation \cite{Cherenkov1937}.\\
Since their physical origin is different, incoherent and coherent CL carry complementary information about the sample. For example in material science, CL spectra are often used to determine characteristic optical resonances and transitions of a given sample. However, if several radiative mechanisms are involved, it is often hard to identify certain spectral properties. Especially, if subtle spectroscopic features of the incoherent CL are overlaid by massive coherent CL background, the qualitative and quantitative analysis becomes difficult, often requiring additional reference measurements. Hence, reliable schemes to separate coherent from incoherent CL are of great interest in many cases. Angle-resolved CL spectroscopy can be used to separate fundamental CL processes by their characteristic angular emission distributions as well as polarization \cite{Brenny2014,Osorio2016}. But, such angle-resolved measurements are not always possible, in particular for transmission electron microscopy (TEM) due to spatial constraints \cite{Yamamoto2016}. In addition, the electron energy-dependence of the CL spectra was recently used to distinguish coherent and incoherent CL in a TEM \cite{Stoger-Pollach2021}.\\
Another way to distinguish CL contributions can be realized by examining the temporal correlations between the photon emission processes. 
Time-resolved coincidence measurements are widely used in physics, in particular in particle physics \cite{Knoll2000}, astrophysics \cite{HamburyBrown1956b} as well as in modern optics and quantum optics \cite{Mandel1984,Kwiat1994,Brendel1999,Eisaman2011,Ecker2021}.
In recent years, time-resolved CL techniques have been introduced to electron microscopy. This allowed to determine the lifetime of sample excitations created by the probing electrons \cite{Merano2005, Tizei2013, Meuret2015, Meuret2018, Feldman2018,Sola-Garcia2021,Meuret2021}. So far, these studies exclusively concentrated on the properties of incoherent CL. In the here presented study, we focus on the temporal correlations of coherent CL, in particular Cherenkov radiation. 

\section{Theory of Cherenkov radiation and coincidence counting}
\begin{figure}[tb]
	\centering
	\includegraphics[width=0.5\textwidth]{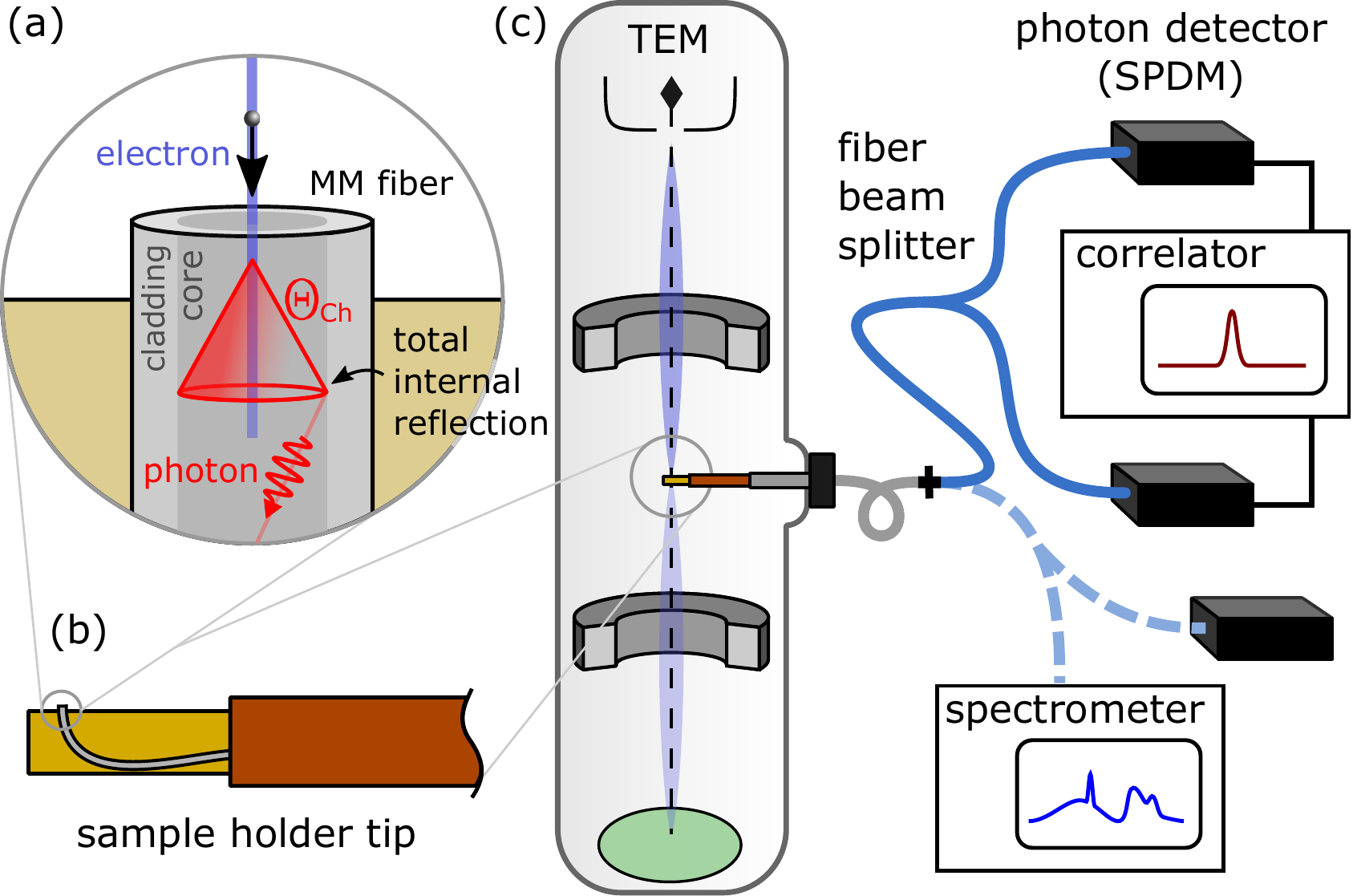}
	\caption{
	CL-TEM setup. (a) Electrons impinge on the core of a multimode (MM) fiber normal to the fiber end face. If the electrons surpass the speed of light in the fiber, they can induce Cherenkov photons on a cone with an angle $\theta_{Ch}$. For emission angles smaller than the critical angle of total internal reflection between the core and the cladding the light will be guided in the fiber. (b) The multimode fiber is bend by 90$^\circ$ and mounted on a custom sample holder tip. (c) The fiber exits the TEM and the output is connected either to a fiber-based Hanbury Brown–Twiss interferometer (see main text) or to an optical spectrometer.
		}
	\label{fig:CR1}
\end{figure} 
Cherenkov radiation is emitted when a charged particle (e.g. an electron) passes through or next to a dielectric medium at a speed greater than the phase velocity of light in that medium. For sufficiently long interaction lengths, the emitted Cherenkov photons are confined to the surface of a cone with vertex angle $\theta_\text{Ch}$ (see Fig.~\ref{fig:CR1}a), defined by \cite{Cherenkov1937}
\begin{equation}
    \cos{\theta_\text{Ch}}=\frac{1}{\beta n}\;,
\end{equation}
where $n$ is the refractive index of the dielectric and $\beta=v/c$ is the velocity of the particle $v$ divided by the speed of light $c$. The photon generation probability $p_\text{Ch}$, per unit wavelength for a given interaction length $L$ and wavelength of the emitted light $\lambda$, can be calculated with \cite{GarciaDeAbajo2010}
\begin{equation}
   \frac{\text{d} p_\text{Ch}}{\text{d}\lambda}= \frac{e^2L}{2\epsilon_0c\hbar\lambda^2}\sin^2\theta_\text{Ch}\;,
    \label{eq:cherenkov_yield}
\end{equation}
where $e$ is the electron charge, $\hbar$ the reduced Planck constant and $\epsilon_0$ the vacuum permittivity. From Eq.~(\ref{eq:cherenkov_yield}) it is evident, that a threshold velocity $v>c/n$ must be exceeded to generate Cherenkov radiation.
\\
Since the generation of CL photons is a stochastic process, there is a finite probability of generating more than one photon per electron. For coherent CL the photons must be generated within a time interval limited by the transit time of the electron $t_\text{trans}=L/v$. For a interaction length of several $\mu$m and typical electron velocities in TEMs ($\sim0.7c$ at 200 kV acceleration voltage), $t_\text{trans}$ is on the order of femtoseconds, thus imposing extremely strong temporal correlations on photons induced by the same electron. For the generation of incoherent CL the situation is different: When electrons irradiate a sample, the probability to stimulate more than one excitation, that will subsequentially lead to incoherent CL, is also non zero. The emission of the photon will, however, occur within the characteristic lifetime $t_\text{life}$ of the excitation. For typical samples,  $t_\text{life}\gg t_\text{trans}$ (e.g. nitrogen vacancies in diamond $\sim 20$~ns \cite{Meuret2015}, defect centers in hexagonal boron nitride $\sim1$~ns \cite{Meuret2015}, or defects in silica $\sim$~5-5000$~\mu$s \cite{Zamoryanskaya2007}). Thus in many settings, the incoherent multi-photon contribution will add to the uncorrelated photon background. Most importantly, the distinct time scales enable the discrimination of coherent from the incoherent CL, when studying the temporal correlations in a coincidence experiment. \\
In the following, let us consider a coincidence measurement of our CL signal using two channels, similar to the setup shown in Fig.~\ref{fig:CR1}c.
In general, the total photon generation probability per electron has a coherent and incoherent CL contribution $p=p_\text{coh}+p_\text{in}$.
The count rate at detector $k\in \{1,2\}$ is then given by 
\begin{equation}
r_k=\alpha_k\, p\, i\;,    
\end{equation}
where $e\!\cdot\! i$ is the electron current and $\alpha_k$ is the single photon detection efficiency of the respective arm, which comprises the collection efficiency, transmission losses and the quantum efficiency of the detector. For uncorrelated events the coincidence rate follows Poissonian statistics and is therefore the same for any delay $\tau$. Given a certain time binning (resolving time) $\Delta t$ the coincidence rate can be approximated by \cite{Knoll2000}
\begin{align}
\label{eq:rtau}
R_{\tau}\approx&\alpha_1\alpha_2\, p^2\, i^2\,\Delta t=\\
&\alpha_1\alpha_2\, (p_\text{coh}^2+2 p_\text{coh}p_\text{in}+p_\text{in}^2)\, i^2\,\Delta t\;.    \nonumber
\end{align}
We can identify contributions from having two uncorrelated coherent photons (i.e. generated by different electrons), one coherent and one incoherent photon and two incoherent photons.
In contrast, the coincidence rate stemming from two correlated coherent CL photons generated by the same electron only contributes to coincidences at $t=0$ and is given by 
\begin{align}
\label{eq:r0}
R_{0}=&\alpha_1\alpha_2\, p_\text{coh}^2\, i \;.
\end{align}
For a given measurement duration $T$, the number of coincidences can be obtained by multiplying the rates with the duration, i.e. $N_l=R_lT$ for $l\in\{0,\tau\}$.
From Eq.~(\ref{eq:rtau}, \ref{eq:r0}) we directly see that the number of accidental coincidence counts depend linearly on time bin size $\Delta t$ and quadratically on the electron current, while the correlated coincidences are a linear function of the current. \\ 
Photon detection setups exhibit a finite timing resolution. When the time binning is much smaller than this value, the correlated coincidence counts are not restricted to a single time bin. Consequently, the central coincidence peak will broaden and its height will decrease by a shape factor $f$ \cite{Sola-Garcia2021}.
\\
To study the photon statistics of a given light field, the second-order autocorrelation function, $g^{(2)}(\tau)$, which provides information about intensity correlations, is often considered \cite{Mandel1995}. It can be obtained from a coincidence measurement by normalizing the coincidence histogram with respect to the value at very large delay, i.e. $\tau\gg1$. Of particular interest is $g^{(2)}$ for zero time delay, which can be interpreted as the likelihood of having two photons at the same time. For light with Poissonian statistics, such as laser light, $g^{(2)}(\tau) = 1$ for any $\tau$. Consequently, $g^{(2)}(0) >(<) 1$ indicates super-(sub-) Poissonian statistics often referred to as (anti)bunching.
For our setting, the second order correlation function for zero time delay reads
\begin{equation}
    g^{(2)}(0)=1+R_0/(f\;R_\tau)\;.
\label{eq:g2}
\end{equation}

\section{Experimental setup}
In order to demonstrate our proposed method, we generate coherent CL (Cherenkov radiation), as well as, incoherent CL (from point defects) in silica using relativistic electrons in a TEM. The experiments are performed using a FEI Tecnai G$^2$ 20 TEM, equipped with a LaB6 electron gun, in combination with a custom-made sample holder, that allows to feed through optical fibers. This enables us to efficiently guide photons that are generated by the sample out of the microscope (see Fig.~\ref{fig:CR1}c). There, the temporal correlations of the light are recorded using a fiber-based Hanbury Brown and Twiss (HBT) interferometer. For the HBT setup, the output-fiber is coupled to a fiber beamsplitter (Thorlabs, TM105R5f1B) and connected to two single photon detector modules (SPDM; Laser Components, COUNT T). The signal is then recorded  and correlated using a time tagging unit (Swabian Instruments, Time Tagger Ultra). The detector and the time tagger exhibit $\sim$500~ps and 49~ps timing jitter, respectively. This yields a lower bound for the resolving time of coincidence counts for our detection system of $\approx1$~ns, mainly limited by the jitter of the photon detectors. We reach a background count level below 400 cts/s on each detector by thoroughly covering all light sources and the bare fibers. The fiber-coupled light can also be sent onto an optical spectrometer (Gatan, Vulcan) via a fiber beamsplitter. While one port is connected to the spectrometer, the second one is connected to a single photon detector to monitor the photon count rate during the acquisition of the spectra. The electron current is measured independently with a picoamperemeter at the drift tube of the image filter \cite{Mitchell2015}.

\begin{figure}[tb]
	\centering
	\includegraphics[width=0.5\textwidth]{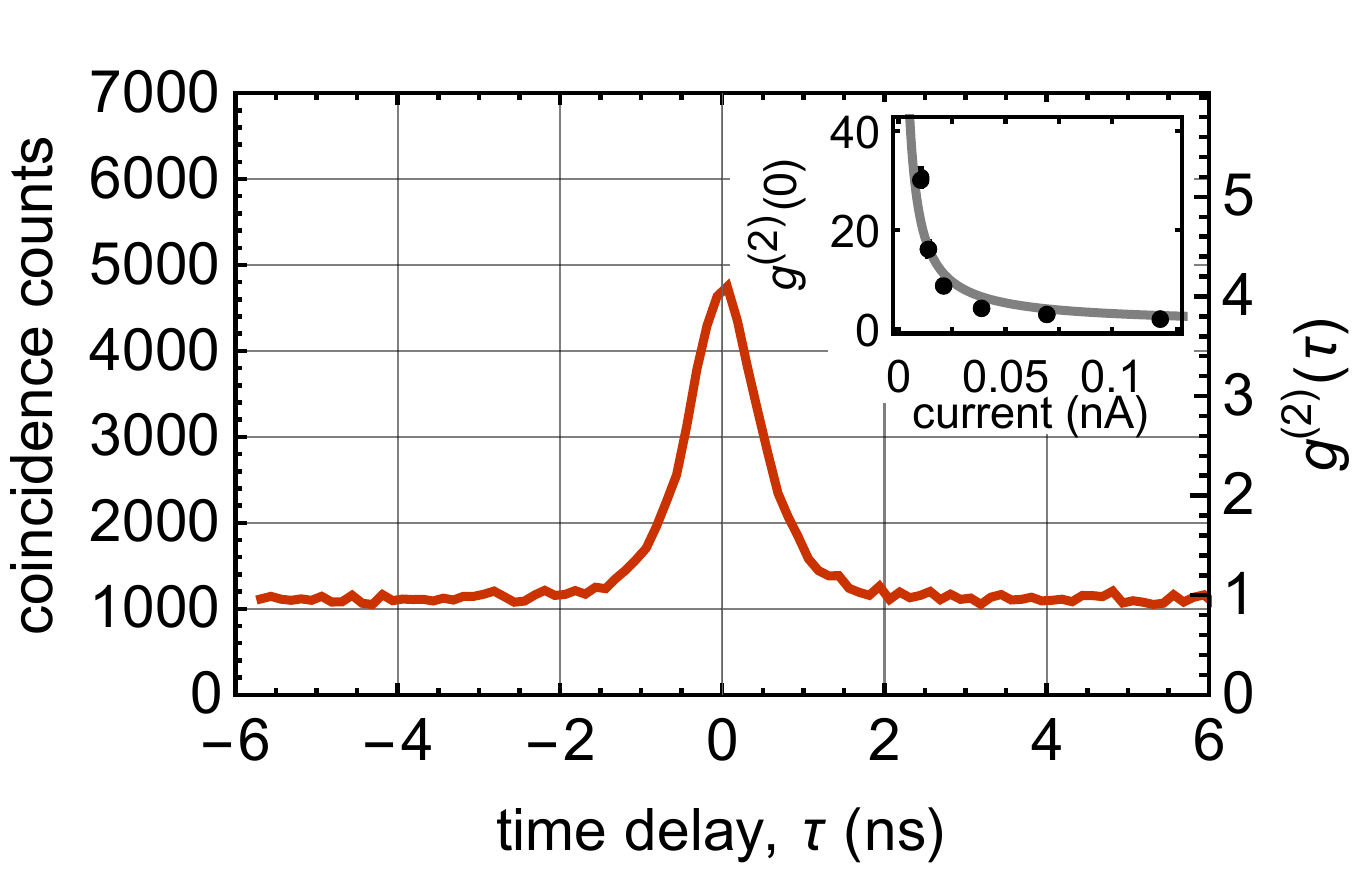}
	\caption{Coincidence counts recorded within 240~s as a function of the time delay between two consecutive photon detections, histogramed with a binning of 125~ps, for a high tension of 200 kV and an electron current of 40 pA. 
The coincidence counts can be translated into the second order correlation function $g^{(2)}$ by normalization. Inset:  $g^{(2)}(0)$ measured as a function of the electron current. The solid line is a quadratic fit according to Eq.~(\ref{eq:g2}).}
	\label{fig:CR2}
\end{figure}

\section{Results}
For our experimental demonstration, we directly induce the light inside the optical fiber. To maximize the Cherenkov yield, we use a core doped multimode fiber (Fiberware, G 90/125 AC), that has a higher refractive core index compared to conventional fibers with a pure silica core. The relatively large NA=0.29 of this fiber allows most of the Cherenkov cone to be collected for acceleration voltages up to 200~kV for a primary electron beam impinging along the fiber axis. To enable this geometry given the spatial constraints (see Fig.~\ref{fig:CR1}b), the fiber is bent using a jet flame torch after cleaving and glued to the tip of the sample holder using a conductive resin (Allied, EpoxiBond 110). To avoid charging, the fiber-tip assembly is coated with $\sim25$~nm of carbon, using a DC magnetron sputter coater (Quroum, Q150T). The electron beam is then focused onto the end face of the fiber and the collected light is guided out of the TEM via a fiber feed-through and analyzed. \\
For a high tension of 200~kV  and an electron current of 40~pA, we obtain an average count rate of 0.39~Mcts/s. This corresponds to an average of $16\times 10^{-4}$ photons that are detected per electron. In Fig.~\ref{fig:CR2}, the recorded coincidence counts are shown for a time binning of 125~ps and a measurement duration of 240~s. A clear peak is visible for zero time delay. The width of the peak (FWHM $\sim~1.1$~ns) is in good agreement with the expected value given the timing jitter of the detection unit. The second order correlation function $g^{(2)}(\tau)$ is computed by normalizing the coincidence counts by the value at $\tau\gg1$. The height of the peak corresponds to $g^{(2)}(0)=4.1\pm 0.1$, indicating clear photon bunching. By in-/decreasing the electron current, and thereby de-/increasing the time between two consecutive electrons, we can further in-/decrease the background count level, thus de-/increase the bunching. As shown in the inset of Fig.~\ref{fig:CR2}, for 10~pA we observed bunching as high as $g^{(2)}(0)=30.6\pm1.6$. This is comparable to previously demonstrated values for incoherent CL \cite{Feldman2018}. Furthermore, we can use the coincidence counts to estimate the ratio of coherent and incoherent CL. For this purpose, we add all coincidence counts within the bunching peak, i.e. $\pm$1.25~ns around $\tau=0$,  $C_0$, and subtract the number of coincidences within the same time interval at large time delays, $C_\tau$. We divide this by the measurement duration and the electron rate $i$ to obtain the coherent photon detection probability $\alpha p_\text{coh}=\sqrt{(C_0-C_\tau)/(i\,T)}$. From $N_\tau$ we can compute the total photon detection probability $\alpha p=\sqrt{C_\tau/(i^2T \Delta t )}$, which allows us to calculate the incoherent contribution $\alpha p_\text{incoh}$ as well. The resulting ratio between coherent and incoherent CL is $p_\text{coh}/p_\text{in}=13\pm2$.

\begin{figure}[tb]
	\centering
	\includegraphics[width=0.5\textwidth]{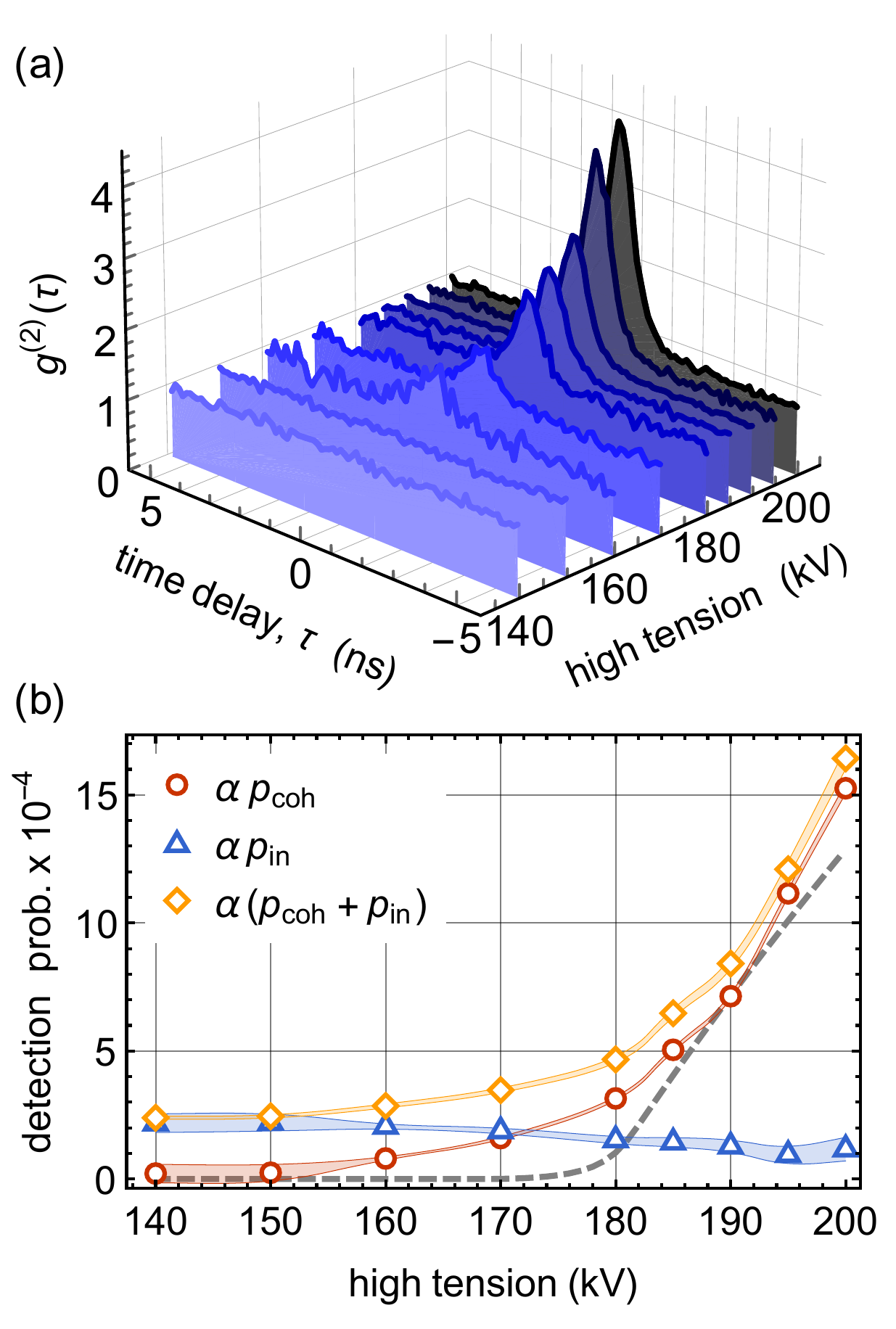}
	\caption{a) Second order correlation function $g^{(2)}(\tau)$ for different settings of the acceleration voltage. b) Coherent, incoherent and total photon detection probability, $\alpha p_\text{coh}$, $\alpha p_\text{in}$ and $\alpha (p_\text{coh}+p_\text{in})$, respectively, as a function of the applied high tension. The shaded area corresponds to the 1$\sigma$ error (statistical and systematic). The dashed line is the result of a theoretical calculation, taking into account the wavelength dependence of the refractive index and the quantum efficiency, that is scaled to approximately fit the measurement.}
	\label{fig:CR3}
\end{figure}

\subsubsection{High tension scan}
As can be seen from Eq.~(\ref{eq:cherenkov_yield}), the Cherenkov yield, at a given refractive index, has a characteristic threshold electron velocity \cite{Stoger-Pollach2008}. We can use this to proof that the observed bunching peak stems from Cherenkov radiation. For this purpose, we conduct the measurement for different acceleration voltages in a range between 140~kV and 200~kV. For the measurement the electron current was varied between 30~pA and 200~pA. The obtained $g^{(2)}(\tau)$ are shown in Fig.~\ref{fig:CR3}a. The emergence of the peak for rising high tension is clearly visible for $U>150$~kV. As before, we can now compute the contribution of coherent and incoherent CL. In Fig.~\ref{fig:CR3}b, the detection probabilities $\alpha p_{coh}$ and $\alpha p_{in}$ are plotted as a function of the applied high tension. As expected for our system, we can exclude coherent CL (Cherenkov radiation) for low acceleration voltages. At approximately 170~kV we detect as much coherent as incoherent CL. The coherent contribution is in good qualitative agreement with the theoretically predicted behavior when accounting for the wavelength dependence of both the refractive index and the quantum efficiency of the photon detectors. Thus, we are confident that the bunching stems from Cherenkov photons and that temporal correlations can be used to separate coherent from incoherent CL.

\subsubsection{Spectral response}

\begin{figure}[tb]
	\centering
	\includegraphics[width=0.45\textwidth]{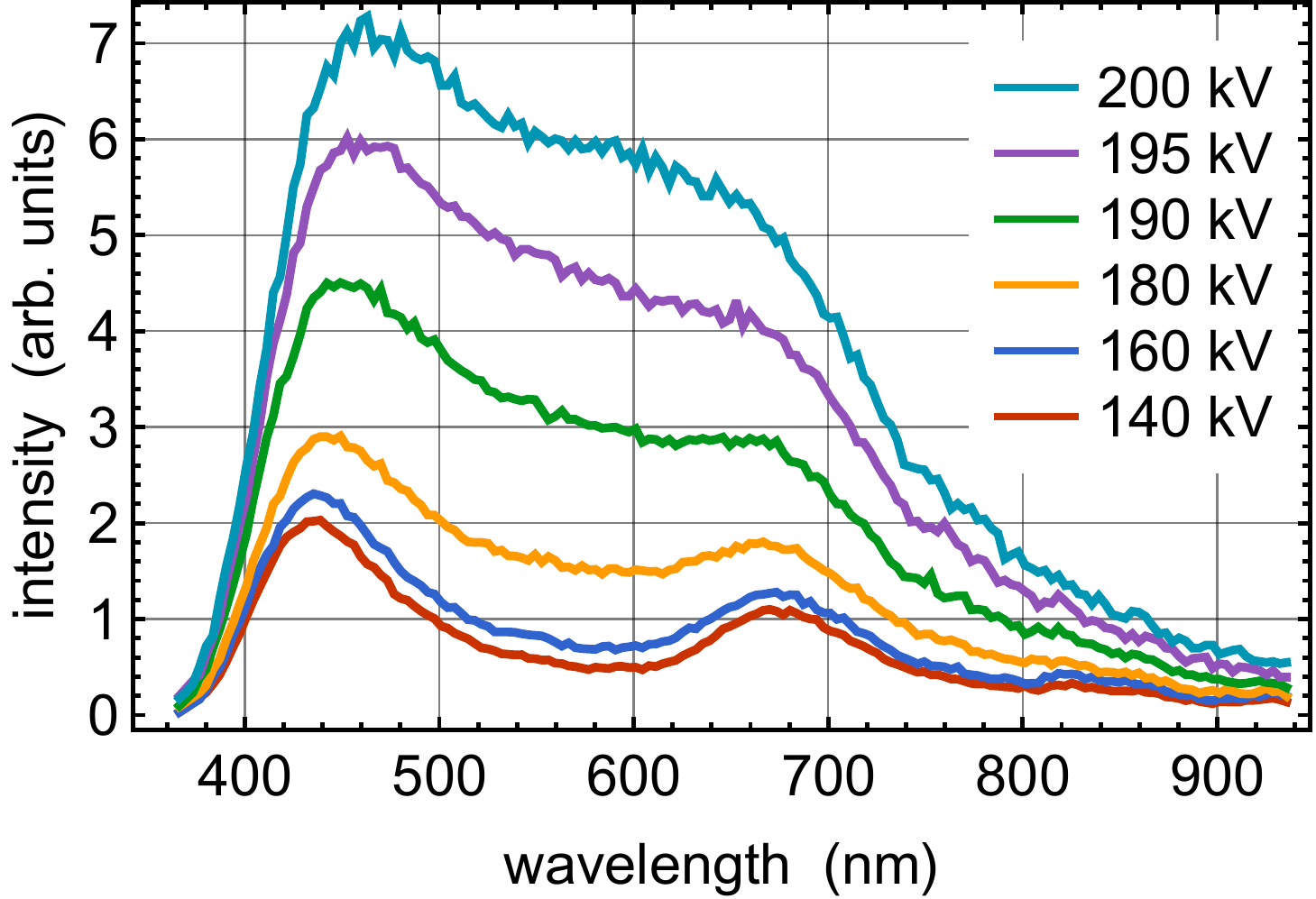}
	\caption{Spectra of the CL induced in the optical multimode fiber core for different acceleration voltages, normalized by the measurement duration and electron current. }
	\label{fig:CR4}
\end{figure}
In addition to the temporal characteristics, coherent and incoherent CL also exhibit specific spectral features, that can in principle be used for discrimination \cite{So2014,Stoger-Pollach2017,Stoger-Pollach2021}. We measure the optical spectrum for different values of high tension. In order to record low-noise spectra, we increase the electron current and the integration time for low acceleration voltages. Fig.~\ref{fig:CR4} shows the raw spectra, only normalized by the integration time and electron current. For small acceleration voltages, two characteristic peaks are visible at $\sim$440~nm  and $\sim$675~nm , that stem from point defects in silica \cite{Zamoryanskaya2007,Kalceff2009}. For larger acceleration voltages, the two peaks are still visible but are superimposed by a broadband background, which qualitatively agrees with the expected $1/\lambda^2$ wavelength dependence of Cherenkov photons. However, a direct comparison of the spectra is not possible since the peak position slightly changes with the applied high tension. This can be attributed to the varying energy deposition within the radiated sample at different settings \cite{Stoger-Pollach2021}.
\section{Discussion and outlook}

To identify coherent CL, the bunching peak has to be much larger than the noise of the uncorrelated background. The resulting signal-to-noise ratio (SNR) scales with $\text{SNR}\propto \sqrt{T/\Delta t}$. Assuming a fixed radiation dose, even rare two photon events (coincidences), originating from coherent CL, can be singled out at low electron flux. This might be beneficial for samples that are prone to radiation damage. Furthermore, we can increase the SNR by using detectors with smaller timing jitter (detectors with 40 ps are commercially available).
It is important to note that the demonstrated method is not restricted to CL generated directly in a fiber, but is also applicable to other CL extraction systems (e.g. via parabolic mirrors). Furthermore, the strong temporal correlations are in principle present for all coherent optical excitations.
In future experiments, the demonstrated coincidence measurement can be employed in a spectrally resolved manner, e.g. by using a gated optical spectrometer. This way, it should be possible to clearly identify the characteristic emission peaks even for large acceleration voltages, despite large background.
In addition, the temporal correlations of the coherent CL, can be used to herald coherent photons for further coherent manipulation or investigation. Thus, our findings can be of use for studying the quantum-coherence properties of cathodoluminescence \cite{Polman2019,Christopher2020}.


\section*{Funding Information}
We acknowledge financial support by the ESQ (Erwin Schrödinger Center for Quantum Science \& Technology) Discovery programme 2020 “CEEP”, hosted by the Austrian Academy of Sciences (ÖAW). P.H., M.Sch. and T.Sp. thank the Austrian Science Fund (FWF) No. Y1121. T.S. acknowledges the financial support of the Austrian Science Fund (FWF), project P29687-N36.

\section*{Acknowledgments}

The authors thank P. Schattschneider and S. L\"offler for fruitful discussions and J. Lazarte Huiza for assistance in early stages of the experiment.

\bibliography{bib}

\appendix*

\end{document}